\documentclass[12pt,a4paper]{article}
\usepackage{amsmath,amssymb,bm,ascmac,bbm,mathtools}
\usepackage[dvipdfmx, usenames]{color}
\usepackage[dvipdfmx]{graphicx}
\usepackage{color}
\usepackage{here}
\usepackage{authblk}

\newcommand{\tdif}[2]{\frac{d#1}{d#2}}

\newcommand{\ddif}[2]{\frac{\delta #1}{\delta #2}}

\newcommand{\mcal}{\mathcal}

\newcommand{\la}{\langle}
\newcommand{\ra}{\rangle}
\newcommand{\End}{\text{End}}

\newcommand{\nn}{\nonumber}
\newcommand{\mn}{{\mu\nu}}

\newcommand{\vep}{\varepsilon}

\begin{document}
\title{Anomalous transport independent of gauge fields}
\author{Mamiya Kawaguchi\footnote{kawaguchi@fudan.edu.cn}} 
\author{Ken Kikuchi\footnote{isjkenkikuchi@gmail.com}}
\affil{Department of Physics and Center for Field Theory and Particle Physics,
Fudan University, 200433 Shanghai, China}
\maketitle

\begin{abstract}
We show that three-dimensional trace anomalies lead to new universal anomalous transport effects on a conformally flat space-time with background scalar fields. 
In contrast to conventional anomalous transports in quantum chromodynamics or quantum electrodynamics, our current is independent of background gauge fields. Therefore, our anomalous transport survives even in the absence of vectorlike external sources. By manipulating background fields, we suggest a setup to detect our anomalous transport. If one turns on scalar couplings in a finite interval and considers a conformal factor depending just on (conformal) time, we find anomalous transport localized at the interfaces of the interval flows perpendicularly to the interval. The magnitude of the currents is the same on the two interfaces but with opposite directions. Without the assumption on scalar couplings, and only assuming the conformal factor depending solely on (conformal) time as usually done in cosmology, one also finds the three-dimensional Hubble parameter naturally appears in our current.
\end{abstract}



$Introduction.-$ 
Quantum field theories (QFTs) often have anomalies in global symmetries, called 't Hooft anomalies \cite{tH80}, which characterize QFTs. 
The global symmetries can be either internal or space-time symmetries. In the latter case, one can still employ the anomalies to study QFTs.
A remarkable example is the $a$ theorem \cite{athm}. In those papers, they coupled a theory to a background metric specified by a ``dilaton." 
Some terms of the dilaton effective action survive even after taking the flat limit, which were used to show the $a$ theorem in four-dimensional QFTs defined on the Minkowski space. An important lesson one can learn from this example is that anomalies can have remnants even after turning off background fields introduced to diagnose anomalies.

Anomalies also lead to interesting phenomena by inducing currents. 
Particularly, the anomalous transport is generated by the chiral anomaly in hot and/or dense environments.
For example, the chiral separation effect\,\cite{Son:2004tq,Metlitski:2005pr} and the chiral magnetic effect\,\cite{Kharzeev:2007jp,Fukushima:2008xe} induce the axial current ${\bm j}_A$ and the vector current ${\bm j}_V$ parallel to an external magnetic field ${\bm B}$ in the dense medium $\mu_V\neq 0$ and the chiral imbalance medium $\mu_A\neq0$, respectively: 
\begin{eqnarray}
\la{\bm  j}_A\ra=\frac{\mu_V}{2\pi^2}e{\bm B},\;\;\;
 \la{\bm j}_V\ra=\frac{\mu_A}{2\pi^2}e{\bm B},
 \label{CME}
 \end{eqnarray}
where $\mu_V$ and $\mu_A$ are the ordinary and chiral chemical potentials, respectively.
Furthermore, in rotating systems, an additional term shows up in the expression of the axial and vector current, which is called the chiral vortical effect\,\cite{Vilenkin:1979ui,Vilenkin:1980zv,Son:2009tf},
\begin{equation}
    \la{\bm  j}_A\ra=\left(\frac{T^2}{6}+\frac{\mu_V^2+\mu_A^2}{2\pi^2}\right){\bm \Omega},\;\;\;
 \la{\bm j}_V\ra=\frac{\mu_V\mu_A}{\pi^2}{\bm \Omega},
 \label{CVE}
\end{equation}
where ${\bm \Omega}$ denotes the angular velocity. This is not the usual background gauge field, but a vectorlike external source, which is not dynamical but rather specified by hand.
Besides the chiral anomaly, the anomalous breaking of the scale symmetry also leads to an anomalous transport. As was discussed in \cite{C16}, an electric current in a curved space-time is generated by the scale electric effect and the scale magnetic effect
\begin{equation}
    \la{\bm j}_V\ra=
    -\frac{2\beta(e)}{e}\frac{\partial \tau(x)}{\partial t}{\bm E}(x)
    +\frac{2\beta(e)}{e}{\bm\nabla}\tau(x)\times {\bm B}(x),
    \label{SEE}
\end{equation}
where $\tau(x)$  represents the local scale factor and $\beta(e)$ is the beta function of quantum electrodynamics (QED).

The induced anomalous current at zero temperature completely relies on the background gauge fields as was shown in Eqs. (\ref{CME})-(\ref{SEE}). Therefore, as background gauge fields are turned off, the conventional anomalous currents vanish (at zero temperature). [The chiral vortical effect (\ref{CVE}) at nonzero temperature also vanishes if one turns off the vectorlike external source $\bm\Omega$.] So, as in the proof of the $a$ theorem, if one is eventually interested in $A=0$ (and $\bm\Omega=0$), all known anomalous transports become trivial. In a sense, this conclusion is natural because a current is a space-time vector, and it seems we need a ``seed'' to produce such a vector-valued quantity. Of course, if one looks at broader currents, one can have nonzero currents in the absence of vectorlike external sources as in scalar QED. (In that case, one has $j_\mu\sim\phi^*\partial_\mu\phi-\phi\partial_\mu\phi^*$ with complex dynamical scalar field $\phi$.) However, if one restricts oneself to currents induced by anomalies, all known examples vanish once vectorlike external sources are turned off. [The conclusion is unaffected even if one takes mass terms into account in (four-dimensional) quantum chromodynamics (QCD) and QED. See Ref. \cite{QCDQED} below.] So it is natural to ask ``is it really necessary to keep background gauge fields (or more precisely, vectorlike external sources) to have nonzero anomalous transports?" We would like to address this question in this paper. And of course the answer is no. One does not have to keep nonzero background gauge fields to have nonzero anomalous transports. We construct an example.

$Setup.-$ 
To concretely derive the anomalous transport independent of external gauge fields, we begin by introducing three-dimensional \cite{3dreason} conformal field theories (CFTs) with a global symmetry $G$ (charge conjugation is not necessarily assumed) on a conformally flat space-time
\begin{equation}
    \gamma_{\mu\nu}(x)=e^{2\tau(x)}\eta_{\mu\nu},\label{gamma}
\end{equation}
where $\mu=0,1,2$, $\eta_\mn=\text{diag}(+1,-1,-1)$ is the Lorentzian metric, and $|\tau|\ll1$. Then via a conformal map, one can translate results on the geometry (\ref{gamma}) to those on the flat space-time. 
On considering renormalization group on such a curved space-time, one has to use local renormalization group (LRG) \cite{LRG,LRGreview}. 
The LRG transformations are realized by Weyl transformations $\gamma_\mn(x)\mapsto e^{2\sigma(x)}\gamma_\mn(x).$
To keep effective action invariant under the transformations, one has to promote all coupling constants $\lambda^I$ to background scalar fields $\lambda^I(x)$, and modify them appropriately according to beta functionals. After introducing the background fields, the partition function $Z$ becomes a functional of background fields \cite{seagull}
\begin{align}
    Z[\tau,\lambda,A]&:=\int\mcal DX\exp\Big\{iS_\text{CFT}[X]+i\int d^3x\tau(x)T^\mu_{\;\;\mu}(x)+i\int d^3x\lambda^I(x)O_I(x)-i\int d^3xA^a_\mu(x)j^{a\mu}(x)\nn\\
    &\hspace{400pt}+\cdots\Big\},\label{gene}
\end{align}
where $X$ collectively denotes dynamical fields,  
$A^a_\mu(x)$ is the background $G$ gauge field, and
$\lambda^I(x)$ denote the coupling constants promoted to the space-time-dependent background scalar fields. The background scalar fields belong to representations $R_I:G\to\End(V^{\dim{R_I}}_{R_I(G)})$ of $G$. 
One then imposes the generating functional be invariant under LRG operator $\Delta_\sigma:=\int d^3x\sigma(x)\{2\gamma_\mn(x)\ddif{}{\gamma_\mn(x)}+\beta^I[\lambda(x)]\ddif{}{\lambda^I(x)}+\beta^a_\mu[\lambda(x)]\ddif{}{A^a_\mu(x)}\}$
up to anomalies, where $\beta^I[\lambda]$ and $\beta^a_\mu[\lambda]\equiv D_\mu\lambda^I\rho^a_I[\lambda]$ are scalar and vector beta functionals, respectively. Namely, one requires $\Delta_\sigma Z[\tau,\lambda,A]=0+(\text{anomaly})$ corresponding to the invariance of observables under a change of an artificial mass scale $\mu$, e.g., $\tdif{Z}{\ln\mu}=0$ in the usual RG.

By using the generating functional (\ref{gene}), one can compute expectation values of operators including trace of the energy-momentum tensor $T^\mu_{\;\;\mu}(x)$, composite (scalar) operators $O_I(x)$, and currents $j^{a\mu}(x)$ by taking functional derivatives with respect to appropriate background fields:
\begin{equation}
    \langle T^\mu_{\;\;\mu}(x) \rangle=-i\frac{\delta\ln Z[\tau,\lambda,A]}{\delta \tau(x)},\quad\la O_I(x)\ra=-i\ddif{\ln Z[\tau,\lambda,A]}{\lambda^I(x)},\quad\langle j^{a\mu}(x)\rangle=i\frac{\delta\ln Z[\tau,\lambda,A]}{\delta A_\mu^a(x)}.\label{vev}
\end{equation}

In CFTs, trace of the energy-momentum tensor vanishes as an operator, however, can suffer from anomalies \cite{trace}. In three dimensions, the general form of the trace anomaly is given by \cite{3dtrace}
\begin{equation}
    \langle T^\mu_{\;\;\mu}\rangle=\epsilon^{\mu\nu\rho}C_{IJK}[\lambda]D_\mu\lambda^I D_\nu\lambda^J D_\rho\lambda^K+\epsilon^{\mu\nu\rho}C^a_I[\lambda] F^a_{\mu\nu}D_\rho\lambda^I,\label{trace}
\end{equation}
where $\epsilon^{\mu\nu\rho}$ is the Levi-Civita tensor with $\epsilon_{012}=+1$, $F^a_{\mu\nu}$ is the field strength of $A$, i.e., $F=dA+[A,A]$, and the covariant derivative is defined by
$D_\mu \lambda^I=[\delta^I_J\partial_\mu+iA^a_\mu (T^a)^I_J]\lambda^J$
with $T^a$ the representation matrices of $R_I$. Since $\epsilon^{\mu\nu\rho}$ is completely antisymmetric, $C_{IJK}[\lambda]$ is also a completely antisymmetric three-tensor. Once a CFT is specified, the coefficient functionals $C_{IJK}$ and $C_I^a$ are fixed. These are analogs of central charges, 
which in general depend on couplings. Similarly, ``central charges'' $C_{IJK}$ and $C^a_I$ depend on couplings in general. Since one promotes coupling constants to background (scalar) fields in LRG, the central charges are functionals of the background scalar fields in the LRG framework. The functionals are subject to constraints \cite{3dtrace} $3\beta^I[\lambda]C_{IJK}[\lambda]+\rho^a_I[\lambda]C^a_J[\lambda]-\rho^a_J[\lambda]C^a_I[\lambda]=0,\,\beta^I[\lambda]C^a_I[\lambda]=0,$ originating from the Wess-Zumino consistency condition \cite{WZ}.

The possibility of three-dimensional trace anomaly was studied holographically \cite{KHS}, and concluded that the anomaly does not exist in odd dimensions if gravity duals of the CFTs exist. However, the trace anomaly has not been ruled out if CFTs do not admit holographic duals. So in the first part of this paper where we impose conformal symmetry, we simply assume an existence of the three-dimensional trace anomaly, and later we relax the assumption of conformal symmetry in which case we do have an example of nonzero trace ``anomaly.''

The current $\langle j^{a\mu}\rangle$ induced by the weak background gauge field $A^a_\mu(x)$, the small local dilatations of the metric $\tau(x)$, and the small space-time dependent coupling constants $\lambda^I(x)$ can be expanded in series of these background fields:
\begin{equation}
    \langle j^{a\mu}\rangle
    =\langle j^{a\mu}\rangle_\tau+\langle j^{a\mu}\rangle_\lambda+\langle j^{a\mu}\rangle_A
    +\langle j^{a\mu}\rangle_{\tau\tau}+\langle j^{a\mu}\rangle_{\tau\lambda}+\langle j^{a\mu}\rangle_{\tau A}
    +\langle j^{a\mu}\rangle_{\lambda\lambda}+\langle j^{a\mu}\rangle_{\lambda A}
    +\langle j^{a\mu}\rangle_{AA}+\cdots.\label{jamu}
\end{equation}
To perform this expansion systematically, we assume the background fields $\tau,\ \lambda,$ and $A$ are all proportional to a small (positive) number $\varepsilon$, $|\varepsilon|\ll1$. The first term $\langle j^{a\mu}\rangle_\tau$ corresponds to the linear response of the current to the operator $\tau(x)T^\mu_{\;\;\mu}(x)$,
\begin{eqnarray}
\langle j^{a\mu}(x)\rangle_\tau=i\int d^3y \langle j^{a\mu}(x)T^\nu_{\;\;\nu}(y)\rangle_0 \tau(y),
\end{eqnarray}
where $\la\cdots\ra_0$ indicates that the expectation value is evaluated by setting background fields to zero after the computation. The correlation function can be calculated easily by taking a functional derivative of (\ref{trace}) with respect to the background gauge field: $\la j^{a\mu}(x)T^\nu_{\:\:\nu}(y)\ra_0=i\ddif{\la T^\nu_{\:\:\nu}(y)\ra}{A^a_\mu(x)}\Big|_{\tau,\lambda,A\to0}=0.$
Thus the first term of (\ref{jamu}) vanishes: $\la j^{a\mu}\ra_\tau=0$. The other terms can be computed in the same way. Generically nonvanishing terms are given by $\la j^{a\mu}\ra_\lambda,\,\la j^{a\mu}(x)\ra_A,$ $\la j^{a\mu}(x)\ra_{\lambda\lambda},\,\la j^{a\mu}(x)\ra_{\lambda A},\,\la j^{a\mu}(x)\ra_{AA},$
and
\begin{align}
    \la j^{a\mu}(x)\ra_{ \tau\lambda}&=
    -\int d^3y\int d^3z\la j^{a\mu}(x)T^\nu_{\;\;\nu}(y)O_I(z)\ra_0\tau(y)\lambda^I(z)\nn\\
    &=-\int d^3y\int d^3z\frac{\delta^2\la T^\nu_{\;\;\nu}(y)\ra}{\delta A^a_\mu(x)\delta \lambda^I(z)}\Biggl |_{\tau,\lambda,A\to0}\tau(y)\lambda^I(z)\nn\\
   &=2\epsilon^{\mu\nu\rho}\partial_\nu \lambda^I(x)
    \partial_\rho\{C_I^a[\lambda(x)]\tau(x)\}\label{taulambda}
\end{align}
These are our main results \cite{confsym}. [Throughout this paper, we consider generic $G$, but if one considers $G=U(1)$, one only has to drop adjoint indices $a,b,\ldots =1,2,\ldots,\dim G$ from our formulas.] As we mentioned in the Introduction, anomalous currents~(\ref{taulambda}) remain finite even after turning off the background gauge field, $A=0$. One may ask how nontrivial this fact is. Actually, as one can show, one cannot obtain nonzero anomalous transports in (four-dimensional) QCD or QED at $A=0$ \cite{QCDQED}. 
Therefore, once vectorlike external sources are switched off, observed anomalous transports can be identified with our current almost uniquely.
Remarkably, the anomalous current with this striking property allows us to have a chance to detect the trance anomaly~(\ref{trace}), as explained below.

Since expectation values of currents may include $\lambda$'s (possibly with covariant derivatives), generically nonzero correlators listed above (\ref{taulambda}) are interesting to study on their own. However, since we would like to focus on the anomalous transport induced by the trace anomaly in this paper, we just consider (\ref{taulambda}) from now on. Note that $C^a_I$ depends on a theory (namely, if one specifies a CFT, the functional is determined), but the form of the current is universal among CFTs with the trace anomaly (\ref{trace}). 
Said differently again, the conventional anomalous transport goes away with vanishing vectorlike external sources $A=0$, so that it would strongly signal an existence of the trace anomaly (\ref{trace}) if one finds this sort of current.
Note that the local counterterm, $\int d^3x \tau(x) \lambda^I(x)J_I(x)$, induces a current similar to our result (\ref{taulambda}). However, the induced current has order ${ O}(\varepsilon^3)$ if $J_I$ depends on $A$. Thus up to $ O(\varepsilon^2)$ the current induced by the local counterterm can be dropped. Therefore, our result is intact up to ${ O}(\varepsilon^2)$ even if the local counterterm is included.   

$Detecting\ our\ anomalous\ current.-$
Now, since we found our anomalous current (\ref{taulambda}) is physical, we would like to discuss its consequences and how to detect it. All observables one has in CFTs are correlation functions because there is no asymptotic state. Therefore we should study correlators including the current operator $j^{a\mu}$. Then we find anomalous current manifests in contact terms of higher point functions of $j^{a\mu}$ and scalar operators $O_I$ or trace of the energy-momentum operator $T^\mu_{\;\;\mu}$. For example, the anomalous current leads to a contact term $\la j^{a\mu}(x)T^\nu_{\;\;\nu}(y)\ra\ni-2i\epsilon^{\mn\rho}\partial_\nu\lambda^I(x)\partial^x_\rho\{C^a_I[\lambda(x)]\delta^{(3)}(x-y)\}$. 
In a realistic experiment, it is difficult to respect the exact conformal symmetry (due to finite size of materials, for example). So as long as we assume conformal symmetry, the only hope to detect the contact term is in lattice simulations. In principle, we believe that the contact term can be detected in numerical lattice simulations. However, since contact terms are sensitive to regularizations, it would be difficult to detect the contact term in lattice simulations. Therefore, in order to discuss a way to detect our current in a more realistic experiment, we would like to relax the assumption of the conformal symmetry.

Without conformal symmetry, one can no longer bring results back to (conformally) flat space-times. However, as we mentioned in Ref. \cite{confsym}, if the current (\ref{taulambda}) cannot be shifted with local counterterms, it is still physical. Therefore,
the remnant of the ``anomalous" current \cite{traceanom} can be observed in experiments. Furthermore, because of smaller space-time symmetry, we do have a theory \cite{KHS} with nonzero trace anomaly (\ref{trace}). The nonzero central charge $C^a_I$ basically corresponds to the anomalous dimension $\gamma^a_{\;\;b}=-\rho^c_I[\lambda]\delta_{bc}(iT^a\lambda)^I$ of the current operator $J^{a\mu}$, where we used the operator equation $D_\mu J^{a\mu}=-(iT^a\lambda)^IO_I$ [see Eq. (43) of the paper]. 
The holography (assumed in the paper) is nowadays applied to condensed matter physics under the name AdS/CMT correspondence, and there also exist some experimental checks (see, say, \cite{KST} for a review). Therefore employing the experimental setup, we believe our anomalous current is detectable in real materials at least in theories studied in \cite{KHS}.

How do the anomalous currents manifest in experiments? Let us discuss the consequences, and a way to detect them.
We propose two places where our anomalous current gives physical consequences; one is real materials and another is the boundary of our observable universe. (``Confinement" of degrees of freedom to such boundaries can be realized by some known mechanisms such as anomaly inflow. However, since we take the effective field theoretical approach, we do not care how exactly the situations are realized.) Let us start from a general consideration.
In the case of $G=U(1)$ and assuming an existence of massless photon field $a$, one can couple the anomalous current to the dynamical-photon field embedded in $A$ as $A=a+\bar A$, where we interpret $\bar A$ as a background $U(1)$ gauge field which is sent to zero in our setup.
Then the anomalous current induces the single-photon production.
The amplitude of the single-photon production tagged with the momentum $q$ and the polarization $\epsilon^{(i)}({\bm q})$ 
is obtained from the generating function, 
\begin{eqnarray}
i{\cal M}_{\tau\lambda}(i;{\bm q}) &=&\langle \epsilon^{(i)}({\bm q}) |0\rangle\nonumber\\
&=&-i
\frac{\epsilon^{(i)\mu} ({\bm q})}{\sqrt{(2\pi)^2 2q^0}}
\int d^3x e^{iq\cdot x }
\langle j_\mu(x)\rangle_{\tau \lambda}.
\label{photon_production}
\end{eqnarray}
This amplitude implies that the real photon is surely emitted from the nonzero anomalous transport. 
Interestingly, the photon emission~(\ref{photon_production}) is certainly produced
even in the absence of vectorlike external sources
in contrast to the single-photon production driven by the conventional anomalous transport~\cite{Fukushima:2012fg}.
So if one detects photons in the absence of background gauge fields, the result suggests the photons are emitted from our anomalous current and not others. In particular, by manipulating background fields (possible in experiments with real materials but difficult at the boundary of our observable universe), one can distinguish our current from the others in this way.
Furthermore, if photons are detected in the situation, it suggests an experimental discovery of the trace anomaly, which has not been found theoretically yet.

In order to make the prediction (\ref{photon_production}) more precise, we would like to manipulate background fields.
We take them as we wish 
in accord with defining properties such as $|\tau|\ll1$. (The following discussion can be repeated assuming conformal symmetry.)

As a first example, let us set $\lambda^I(x)=\vep^I\theta(x^1)\theta(l-x^1)$ where $\vep^I$'s are numbers of order $\vep$ and $l$ is some length. This choice is motivated by doping in condensed matter physics; one can dope a material with a fermion $\psi$ only in an interval $x^1\in[0,l]$. Then an example of position-dependent scalar coupling is given by a Yukawa coupling $\lambda\phi\bar\psi\psi$ between $\psi$ and a scalar field $\phi$ with $\lambda\sim\theta(x^1)\theta(l-x^1)$. Preparing more Yukawa couplings $\lambda^I$, we can model the situation as $\lambda^I$'s are turned on only in the interval, $\lambda^I(x)=\vep^I\theta(x^1)\theta(l-x^1)$. Adopting this choice of scalar background fields, we get
\begin{equation}
\begin{split}
    \la j^{a0}(x)\ra_{\tau\lambda}&=2\varepsilon^IC^a_I(\varepsilon/2)\partial_2\tau(x)\left[\delta(x^1)-\delta(l-x^1)\right],\\
    \la j^{a1}(x)\ra_{\tau\lambda}&=0,\\
    \la j^{a2}(x)\ra_{\tau\lambda}&=-2\varepsilon^IC^a_I(\varepsilon/2)\partial_0\tau(x)\left[\delta(x^1)-\delta(l-x^1)\right].
\end{split}\label{taulambda1}
\end{equation}

We would like to make three comments on this result: (i) This result shows that (if one prepares doped materials as we described) our anomalous current, perpendicular to the interval $x^1\in[0,l]$, is generated at the interface of doped intervals. (ii) The form of anomalous current is independent of models (only central charges $C^a_I$ depend on theories), and current necessarily exists if $C^a_I$ is nonzero as in holographic QFTs \cite{KHS}. (iii) Even though the current is proportional to small numbers $\vep^I$, its magnitude is amplified by Dirac's delta, which would enable experimental detection.
Consequently, the anomalous current produces the nonzero amplitude (\ref{photon_production}). It strongly indicates that the real-photon emission is induced at the interfaces of, say, doped materials, which is a significant consequence of the nonzero anomalous current independent of the external gauge fields. 
Next, we shall consider a second example with $\tau(x)=\tau(x^0)$. This form is motivated by cosmology (although we are focusing on three dimensions and not four dimensions); in cosmology, assuming the homogeneous isotropic universe, one usually considers a metric $ds^2=dt^2-a^2(t)\delta_{ij}dx^idx^j,$
where $t$ is the cosmic time, and $a(t)$ expresses the size of the universe. Via a coordinate transformation $dt=a[t(\eta)]d\eta,$
the metric reduces to the conformally flat one: $ds^2=a^2[t(\eta)]\left(d\eta^2-\delta_{ij}dx^idx^j\right)$
($\eta$ is called the conformal time). 
Identifying $\eta=x^0$, we have $\tau(x^0)=\ln a[t(x^0)]$. We cannot apply the setup to the $bulk$ but to the $boundary$ of the observable universe as follows; The (spatial) boundary of our observable universe can be modeled by 
a 2-sphere with radius $r_\text{universe}$. Since its curvature scaling as $1/r_\text{universe}^2$ is negligible, it can be approximated by the flat 2-space. Therefore, the boundary of the observable universe (approximately) has a conformally flat metric, and our setup is applicable. The three-dimensional space-time has the Hubble parameter $H_{3d}(t):=a^{-1}(t)da(t)/dt$, which characterizes the expansion of our universe. The three-dimensional Hubble parameter naturally enters in the derivative of $\tau$:
\begin{equation}
    \tdif{\tau(x^0)}{x^0}=H_{3d}[t(x^0)]a[t(x^0)].\label{dottau}
\end{equation}

Now, let us turn to the current. Employing $\tau=\tau(x^0)$, one obtains formulas; however, its physical meaning is unclear. To draw physical meaning,
we take a reference time $t=t_0$, and set $a(t_0)=1$. Typically, $t_0$ can be taken as the current age of the universe, which is ``long" in the ordinary sense.
We consider a small deviation from the reference time, $|(t-t_0)/t_0|\ll1$.
Then we have $H_{3d}[t(x^0)]=\dot a(t_0)+{ O}[|(t(x^0)-t_0)/t_0|]$. The leading term is the three-dimensional Hubble constant: $\dot a(t_0)=\left(H_{3d}\right)_0$. Expanding our currents in powers of $|(t(x^0)-t_0)/t_0|$, we obtain
\begin{equation}
\begin{split}
\la j^{a0}(x)\ra_{\tau\lambda}&=
{ O}[|(t(x^0)-t_0)/t_0|],\\
\la j^{a1}(x)\ra_{\tau\lambda}&=
2\partial_2 \lambda^I(x)C_I^a[\lambda(x)]\left(H_{3d}\right)_0+{ O}\{|(t(x^0)-t_0)/t_0|\},\\
\la j^{a2}(x)\ra_{\tau\lambda}&=
-
2\partial_1 \lambda^I(x)C_I^a[\lambda(x)]\left(H_{3d}\right)_0+{ O}\{|(t(x^0)-t_0)/t_0|\}.
\end{split}\label{taulambda2}
\end{equation}
This result implies that the anomalous current is possibly triggered by the expansion of the universe, which causes the photon emission from the edge of the observable universe. So, it is expected that the emitted photons mix with other sources of radiation such as cosmic microwave background (CMB). 
To see this, let us consider another set of background fields.


The final example is $\lambda^I(x)=\vep^I\theta(x^1)\theta(l-x^1)$ and $\tau(x)=\tau(x^0)$. Then we immediately obtain
\begin{equation}
\la j^{a0}(x)\ra_{\tau\lambda}=0,\quad\la j^{a1}(x)\ra_{\tau\lambda}=0,\quad\la j^{a2}(x)\ra_{\tau\lambda}=-2\varepsilon^IC^a_I(\varepsilon/2)\tdif{a[t(x^0)]}{t(x^0)}\left[\delta(x^1)-\delta(l-x^1)\right].\label{taulambda3}
\end{equation}
In order to relate this result to cosmology, one can take, for example, particles on the boundary of our observable universe as sources of photon emission. Such particles, modeled by a 3-ball, have one-dimensional (spatial) interfaces when projected to the boundary. Then with suitable modifications on $\lambda^I(x)$, our result (\ref{taulambda3}) predicts currents generated at the interfaces. Accordingly, photons are emitted at the interfaces. The photons mix with CMB, and the 
scattered sources of photons may explain fluctuations in CMB. It is intriguing to push this consideration forward.

\section*{Acknowledgment}
We are grateful to Xu-Guang Huang and Shinya Matsuzaki for useful comments.

\appendix
\setcounter{section}{0}
\renewcommand{\thesection}{\Alph{section}}
\setcounter{equation}{0}
\renewcommand{\theequation}{\Alph{section}.\arabic{equation}}


\begin{thebibliography}{30}
\bibitem{3dreason} One can repeat our analysis below also in other dimensions including $d=4$. The only condition which has to be satisfied in a space-time dimension one is interested in would be that the trace anomaly in that dimension have terms quadratic in background fields so that the anomaly induces an anomalous transport to the second order. Therefore we could perform our analysis in other dimensions but we focus on $d=3$ for several reasons. Firstly, three-dimensional trace anomaly has fewer terms, making the analysis simpler. Secondly, it turns out that we can experimentally address the longstanding problem of an existence of three-dimensional trace anomaly.
\bibitem{seagull} Since we consider responses to variation of background fields up to their quadratic order, ``seagull terms" abbreviated by $(\cdots)$ may also contribute. Here, we abused the terminology ``seagull term," and meant all quadratic terms in background fields not just gauge fields $A$ but also scalar fields $\tau$ and $\lambda$. So they may include not just Chern-Simons terms $\int\text{tr}AdA+\cdots$, but also terms such as $\int d^3x\tau(x)\lambda^I(x)J_I(x)$ with some local operators $J_I$. We discuss whether these terms contribute or not later.
\bibitem{confsym} Actually, the current can be found even when one relaxes the assumption of conformal symmetry. In that case, the right-hand side of (\ref{trace}) also contains contributions from operators, but if the terms consist of just background fields that cannot be tuned away with local counterterms, one can still repeat our computation leading to (\ref{taulambda}). We are assuming conformal symmetry to relate our results on a conformally flat space (\ref{gamma}) to those on flat space-time; however, if one gives up translating the results back to flat space-time, one does not have to assume the symmetry.
\bibitem{QCDQED} In QCD and QED, quantum anomalies of the scale symmetry or the chiral symmetry induce the anomalous transports depending on background gauge fields. If one takes explicit scale/chiral symmetry breaking into account, the nonzero fermion mass also gives additional contributions to the trace of the energy-momentum tensor or divergence of the current, respectively. So, one might deduce that the nonzero fermion mass could induce anomalous transports independent of background gauge fields, which may be derived from a correlation function like $\langle j^\mu(x) \bar\psi \psi(y) \rangle_0$; the one-point function $\la\bar\psi\psi(y)\ra$ can be linear in background gauge fields, resulting in additional contributions to $\la j^\mu(x)\bar\psi\psi(y)\ra_0$ independent of background gauge fields. However, the correlator $\la j^\mu(x)\bar\psi\psi(y)\ra_0$ vanishes in QCD and QED due to their (exact) charge conjugation symmetry. Since all contributions proportional to background gauge fields one obtains by bringing them down from the action vanish in the process $A\to0$, one concludes that in QCD and QED one cannot find nonzero anomalous transports in the absence of background gauge fields.
\bibitem{traceanom} Strictly speaking, since there is no conformal symmetry, (\ref{trace}) is no longer an anomaly. However, we continue to simply call the current induced by (\ref{trace}) as anomalous current following \cite{C16} because we are assuming it is physical.
\end{thebibliography}
\end{document}